\documentclass[%
 amsmath,amssymb,
 aps,
]{revtex4-1}
\usepackage{graphicx}
\usepackage{dcolumn}
\usepackage{bm}
\usepackage{epstopdf}

\begin{document}

\preprint{APS/123-QED}

\title{Solvababe supersymmetric algebraic model for  descriptions of transitional even and odd mass nuclei near the critical point of the spherical to unstable shapes}

\author{ M. A. Jafarizadeh}
\email{jafarizadeh@tabrizu.ac.ir}
\affiliation{Department of Theoretical Physics and Astrophysics,
University of Tabriz, Tabriz 51664, Iran.}
\affiliation{Research Institute for Fundamental Sciences, Tabriz 51664, Iran}

\author{ M.Ghapanvari}
\email{M.ghapanvari@tabrizu.ac.ir}
\affiliation{Department of Nuclear Physics, University of Tabriz, Tabriz 51664, Iran.}
\affiliation{The Plasma Physics and Fusion Research School, Tehran, Iran.}

\author{ N.Fouladi}
\email{fooladi@tabrizu.ac.ir}
\affiliation{Department of Nuclear Physics, University of Tabriz, Tabriz 51664, Iran.}

\date{\today}

\begin{abstract}
Exactly solvable solution for the spherical to gamma - unstable transition in transitional nuclei based on dual algebraic structure and nuclear supersymmetry concept is proposed. The duality relations between the unitary and quasispin algebraic structures for the boson and fermion systems are extended to mixed boson-fermion system.  It is shown that the relation between the even-even and odd-A neighbors implied by nuclear supersymmetry in addition to dynamical symmetry limits can be also used for transitional regions. The experimental evidences are presented for even- even [E(5)] and odd-mass [E(5/4)] nuclei near the critical point symmetry.

\end{abstract}

\maketitle
Exactly solvable models play a key role in understanding some many - body problems in nuclear physics. A model may be solvable if its energy levels can be determined analytically  and eigenstates are identified completely by the quantum numbers of a subgroup chain\cite{1,2}.
There are  some many-body problems in nuclear physics that are exactly solvable, for example, three limits of dynamical symmetry of the interacting boson model\cite{2}.
The IBM describes even-even nuclei in terms of correlated pairs of nucleons with L = 0, 2 treated as bosons (s, d bosons)\cite{3}. The  N-boson system space is spanned by the irrep $ \mathrm{[N]} $ of $ \mathrm{U^{B}(6)} $ \cite{3}. The interacting boson - fermion model also explains odd-A nuclei in terms of correlated pairs,  s and d bosons , and unpaired particles of angular momentum j (j fermions)\cite{4}. The states of the boson - fermion system can be classified according to the irreducible representation  $ \mathrm{[N] \times [1]} $ of $ \mathrm{U^{B}(6) \times U^{F}(M)} $ where M is the dimension of the single particle space. The IBM and IBFM can be unified into a supersymmetry (SUSY) approach that was discovered into nuclear structure physics in the early 1980 \cite{5,6}. The experiments performed at various laboratories have  confirmed the predictions made using a SUSY scheme \cite{7}. Originally, nuclear supersymmetry was considered as symmetry among pairs of nuclei consisting of an even-even and an odd-even nuclei\cite{5,6}. The supersymmetric representations $[\mathcal{N}\}$ of $\mathrm{U(6/ M)}$
spanned a space  explaining the lowest states of an even-even nucleus with $\mathcal{N}$ bosons and an odd-A nucleus with $\mathcal{N}-1$ bosons and an unpaired fermion\cite{4}. The nuclear supersymmetry has been used successfully in description of the dynamical symmetry limits of the even-even and odd-A nuclei\cite{4,5,6}. Virtually simultaneously,  with the introduction of the nuclear supersymmetry, the idea of spherical-deformed phase transitions at low energy in finite nuclei germinated\cite{8,9}. Studies of QPTs in odd-even nuclei with supersymmetric scheme had implicitly been initiated years before by A. Frank et al.\cite{10}. They have studied successfully a combination of $\mathrm{U^{BF}(5)}$ and $\mathrm{SO^{BF}(6)}$ symmetry by using $\mathrm{U(6/12)}$ supersymmetry for the Ru and Rh isotopes. Iachello \cite{11}  extended the concept of critical symmetry to critical supersymmetry and provided a benchmark for the study of shape phase transition in odd-even nuclei and also J.Jolie et al.\cite{12}  studied QPTs in odd-even nuclei using a supersymmetric approach in interacting Boson-Fermion model.

The purpose of this letter is to point out that we have proposed a new solvable model for describing QPT for even and odd mass nuclei by using supersymmetry approach.
In this paper, concept of supersymmetry and phase transitions  are brought together by using the generalized quasi-spin algebra and Richardson - Gaudin method. We consider  the  state of fermions with spin j=3/2 coupled to boson core, however, our method  is  applicable  for the whole systems with other values of spin of the fermions, j. In order to obtaining an algebraic solution for transitional region, we have used of dual algebraic structures. The  duality symmetries are a powerful tool in relating  the Hamiltonians with the number-conserving unitary and number-nonconserving quasi-spin algebras for system with pairing interactions\cite{13,14}. These relations are obtained  for both bosonic
and fermionic systems\cite{13,14}. We have established the duality relations for mixed boson-fermion system.
In this paper, we display that the relation between the even-even and odd-A neighbors implied by nuclear supersymmetry in addition to dynamical-symmetry limits can also be used for transitional regions. So, the testing SUSY in all nuclear regions( dynamical symmetry limits and transitional region ) are possible.
We investigate the change in level structure induced by the phase transition by doing a quantal analysis. The experimental evidences are presented for even- even  [E(5)]  and odd-mass  [E(5/4)] nuclei near the critical point symmetry.

Details of the group theoretical description
will be omitted from this paper and will be given in a
subsequent detailed publication.
The quasi-spin algebras have been explained in detail in Refs \cite{13,15}. The generalized quasi-spin algebra contains both bosonic (B) and fermionic (F) operators defined as \cite{16}
\begin{equation}
S_{BF}^{0}=\frac{1}{2}(n_{b}+n_{f})+ \frac{1}{4}(N-M)
\end{equation}
\begin{eqnarray}
S_{BF}^{+}=\frac{1}{2} \sum\limits_{m}(-1)^{2\mp m}b_{m}^{+}b_{-m}^{+}\mp \frac{1}{2} \sum\limits_{m'}(-1)^{j\mp
 m'}a_{jm'}^{+}a_{j-m'}^{+}=\frac{1}{2}(b^{+}\cdot b^{+} )\mp (a_{j}^{+} \cdot a_{j}^{+} )
\end{eqnarray}
\begin{equation}
S_{BF}^{-}=\frac{1}{2} ( \widetilde{b}.\widetilde{b}  ) \pm \frac{1}{2} ( \widetilde{a}_{j}.\widetilde{a}_{j}  )
\end{equation}
Where $\mathrm{n_{b}}$ and $\mathrm{n_{f}}$are the boson and fermion number operators, respectively.Thus the operators $\mathrm{(S_{BF}^{\pm}}$,$\mathrm{S_{BF}^{0})}$ form a generalization of the usual fermionic and bosonic quasi-spin algebras with commutation relations given as \cite{16}
\begin{equation}
[S_{BF}^{0},S_{BF}^{\pm} ]=\pm S_{BF}^{\pm}  \quad                    ,   \quad [S_{BF}^{+},S_{BF}^{-} ]=-2S_{BF}^{0}
\end{equation}
The bosonic quasi-spin operators satisfy the commutation relations of the quasi-spin SU(1,1) algebra \cite{13,15,16} while the fermionic quasi-spin operators satisfy the commutation relations of the  quasi-spin SU(2) algebra \cite{13,15,16}.
The irreps of generalized quasi-spin algebra (GQA) are given in terms of the eigenvalues of the quadratic Casimir invariant and  the quasi-spin operator $\mathrm{S_{BF}^{0}}$.
The basis states of an irreducible representation (irrep)GQA, $\mathrm{|k,\mu \rangle}$ ,are determined by a single number $ \mathrm{k}$ , susceptible of any positive number and $ \mathrm{\mu=k,k+1,....}$  \cite{16} Therefore,
\begin{equation}
|k,\mu \rangle=|\frac{N-M}{4}+\frac{\nu_{B}+\nu_{F}}{2},\frac{ N-M}{4}+\frac{N_{B}+N_{F}}{2} \rangle
\end{equation}
The basis states are determined  considering the fact that fermionic part of the action of $\mathrm{S_{BF}^{+}} $ is restricted by the Pauli Exclusion Principle and the action of $\mathrm{S_{BF}^{-} }$ terminates when a state of $\mathrm{\nu}$ unpaired particles is reached i.e
$\mathrm{S_{BF}^{-}|\nu_{B},\nu_{F} \rangle =0}$\cite{16}.
In order to investigation the phase transition in atomic nuclei  according to IBM, we have considered two kinds of bosons with L = 0, 2  (s, d bosons)\cite{3}. The generators of $ \mathrm{SU^{d} (1,1)}$ and $ \mathrm{SU^{s} (1,1)}$ are designated by considering and putting s and d operators only in the left part of Eqs.[1-3]and
$ \mathrm{SU^{sd} (1,1)}$ that is the s and d boson pairing algebras generated by \cite{15}
\begin{equation}
 S^{+}(sd)=\frac{1}{2} (d^{+}.d^{+}\pm s^{+^2} ) \quad ,
 \quad S^{-}(sd)=\frac{1}{2} (\widetilde{d}.\widetilde{d} \pm s^{2})
 \quad , S^{0}(sd)=\frac{1}{4}{\sum_{\nu}({d_{\nu}^{+}d_{\nu}+d_{\nu}d_{\nu}^{+}})}
  +\frac{1}{4} (s^{+} s+ss^{+} )
  \end{equation}
Because of the duality relationships \cite{13,14,15}, it is known that
in even - even nuclei the base of $\mathrm{U(5)\supset SO(5)} $ and $\mathrm{ SO(6)\supset SO(5)} $ are
simultaneously the basis of $ \mathrm{SU^{d} (1,1)\supset U(1)} $ and $
\mathrm{SU^{sd} (1,1)\supset U(1)} $, respectively. By the use of duality
relations \cite{13,15}, the Casimir operators of SO(5) and
SO(6) can also be expressed in terms of the Casimir operators
of $ \mathrm{SU^{d} (1,1)}$ and $\mathrm{ SU^{sd} (1,1)}$, respectively
\begin{equation}
  \hat{C } _{2}(SU^{d} (1,1))=\frac{5}{16}+\frac{1}{4} \hat{C } _{2}(SO^{B}(5) )
\end{equation}
\begin{equation}
  \hat{C } _{2}(SU^{sd} (1,1))=\frac{3}{4}+\frac{1}{4} \hat{C } _{2}(SO^{B}(6) )
\end{equation}
The correspondence between the basis vectors in this case was shown in Ref.[15].
For a mixed boson-fermion system, the chain of subalgebras of unitary superalgebras U(6/M ) for j=3/2 is shown in Fig.1 and also two-level pairing system has two dynamical symmetries defined with respect to the generalized quasispin algebras, corresponding to either the upper or lower subalgebra chains in Eq.(9).
\begin{equation}
GQA_{1}^{sf}\otimes GQA_{2}^{df} \supset \left \{
\begin{array}{cl}
GQA_{1,2}^{sdf} \\
U_{1}^{sf}(1)\otimes U_{2}^{df}(1)\\
\end{array}\right\}
\supset U_{1,2}^{sdf}
\end{equation}
The upper subalgebra chain is corresponding to strong-coupling dynamical symmetry limit while lower chain is weak-coupling limit.
Therefore, for odd-A nuclei, we have obtained the dual relation between the Casimir operators $\mathrm{Spin^{BF}(5)}$ and $\mathrm{GQA^{df}}$ (generalized quasispin algebra of d bosons and single fermion with j=3/2) as

If $\tau_{1}=v_{d}-\frac{1}{2} $ and $ \tau_{2}=\frac{1}{2}$
\begin{equation}
\hat{C } _{2}(GQA^{df} )=\frac{1}{4} \hat{C } _{2}(Spin^{BF}(5) )-\frac{1}{4}(\tau_{1}+\frac{3}{4})
\end{equation}
If $\tau_{1}=v_{d}+\frac{1}{2} $ and $ \tau_{2}=\frac{1}{2}$
\begin{equation}
\hat{C } _{2}(GQA^{df} )=\frac{1}{4} \hat{C } _{2}(Spin^{BF}(5) )-\frac{1}{4}(3\tau_{1}+\frac{7}{4})
\end{equation}
By the use of duality
relations, the correspondence between the basis vectors $\mathrm{Spin^{BF}(5)}$ and $\mathrm{GQA^{df}}$ is
\begin{equation}
|\mathcal{N};[N_{B}=N],{N_{F}=1},\nu_{d},(\tau_{1}=v_{d}-\frac{1}{2},\tau_{2}) ,n_\Delta JM \rangle = |\mathcal{N};k^{d}=\frac{1}{2}(\nu_{d}+\frac{5}{2}),k^{df}=\frac{\tau_{1}+2}{2},\mu^{df}=\frac{1}{4}+\frac{1}{2}(n_{d}+n_{f}) ,n_\Delta JM \rangle
\end{equation}
\begin{equation}
|\mathcal{N};[N_{B}=N],{N_{F}=1},\nu_{d},(\tau_{1}=v_{d}+\frac{1}{2},\tau_{2}) ,n_\Delta JM \rangle = |\mathcal{N};k^{d}=\frac{1}{2}(\nu_{d}+\frac{5}{2}),k^{df}=\frac{1+\tau_{1}}{2}, \mu^{df}=\frac{1}{4}+\frac{1}{2}(n_{d}+n_{f}) ,n_\Delta JM \rangle
\end{equation}
The Casimir operator
of $\mathrm{Spin^{BF}(6)}$ and $\mathrm{GQA^{sdf}}$ (generalized quasispin algebra of d and s bosons with fermion j=3/2 ) has the following correspondence

If $\mathrm{\sigma_{1}=\sigma-\frac{1}{2} }$ and $ \mathrm{\sigma_{2}=|\sigma_{3}|=\frac{1}{2}}$
\begin{equation}
\hat{C } _{2}(GQA^{sdf} )=\frac{1}{4} \hat{C } _{2}(Spin^{BF}(6) )-\frac{3}{4}(\sigma_{1}+\frac{3}{4})
\end{equation}
If $\mathrm{\sigma_{1}=\sigma+\frac{1}{2}} $ and $ \mathrm{\sigma_{2}= |\sigma_{3}|=\frac{1}{2}}$
\begin{equation}
\hat{C } _{2}(GQA^{sdf} )=\frac{1}{4} \hat{C } _{2}(Spin^{BF}(6) )-\frac{1}{4}(\sigma_{1}+\frac{3}{2})
\end{equation}
By use of duality
relations, the correspondence between the basis vectors $\mathrm{Spin^{BF}(6)}$ and $\mathrm{GQA^{sdf}}$ is
\begin{equation}
|\mathcal{N};[N_{B}=N],{N_{F}=1},\sigma,(\sigma_{1},\sigma_{2},\sigma_{3}),(\tau_{1},\tau_{2}) ,n_\Delta JM \rangle = |\mathcal{N};k^{sdf}=\frac{1}{2}(\sigma_{1}+\frac{3}{2}),\mu^{sdf}=\frac{1}{4}+\frac{1}{2}(n_{s}+n_{d}+n_{f}) ,n_\Delta JM \rangle
\end{equation}
\begin{equation}
|\mathcal{N};[N_{B}=N],{N_{F}=1},\sigma,(\sigma_{1},\sigma_{2},\sigma_{3}),(\tau_{1},\tau_{2}) ,n_\Delta JM \rangle = |\mathcal{N};k^{sdf}=\frac{1}{2}(\sigma_{1}+\frac{5}{2}),\mu^{sdf}=\frac{1}{4}+\frac{1}{2}(n_{s}+n_{d}+n_{f}) ,n_\Delta JM \rangle
\end{equation}
These relations have been used to effect simplifications of the calculations for two-level and multi-level systems\cite{13}.
The infinite dimensional generalized quasispin algebra is generated by the use of \cite{15}
\begin{equation}
S_{BF,n}^{\pm}=c_{s}^{2n+1} S_{B}^{\pm} (s)+c_{d}^{2n+1} S_{B}^{\pm} (d)+c_{f}^{2n+1} S_{F}^{\pm} (f)
\end{equation}
\begin{equation}
S_{BF,n}^{0}=c_{s}^{2n} S_{B}^{0} (s)+c_{d}^{2n} S_{B}^{0} (d)+c_{f}^{2n} S_{F}^{0} (f)
\end{equation}
Where $\mathrm{c_{s}}$ ,$\mathrm{c_{d}}$ and $\mathrm{c_{f}}$ are real parameters and  n  can be
$\mathrm{0,\pm1,\pm2,....}$. These generators satisfy the commutation
relations
\begin{equation}
[S_{BF,m}^{0}  ,S_{BF,n}^{\pm} ]=\pm S_{BF,m+n}^{\pm}
\quad\quad     ,     \quad\quad         [S_{BF,m}^{+},S_{BF,n}^{-}
]=-2S_{BF,m+n+1}^{0}
\end{equation}
Then,$\mathrm{{S_{BF,m}^{\mu},\mu=0,+,-; m=\pm1,\pm2,...} } $generate an
affine generalized quasispin algebra $ \mathrm{\widehat{GQA}} $ without central extension.
The detailed description of the  $\mathrm{U(6/4)}$  Supersymmetry in U(5) and O(6) limits can be found in \cite{4}. By employing the generators of $ \mathrm{\widehat{GQA}} $ and Casimir operators of subalgebras, the following Hamiltonian for transitional region between U(5)-O(6) limits is prepared
\begin{equation}
\hat{H }=S_{BF,0}^{+} S_{BF,0}^{-}+\alpha S_{BF,1}^{0}+\beta\hat{C }
_{2}(Spin^{BF}(5) )+\gamma\hat{C }
_{2}(spin^{BF} (3) )
\end{equation}
It can be shown that Hamiltonian Eq.(21) is equivalent to a boson Hamiltonian for the even-even nuclei if acting on the $ \mathrm{[N] \times [0]} $ representation of $ \mathrm{U^{B}(6)\times U^{F}(4)}$ and with boson-fermion Hamiltonian for odd-A nuclei if acting on the $ \mathrm{[N] \times [1]} $ representation of $ \mathrm{U^{B}(6)\times U^{F}(4)}$. In odd-A nuclei, Hamiltonian Eq.(21)  is equivalent to $ \mathrm{O^{BF}(6)}$
Hamiltonian when $\mathrm{c_{s}=c_{d}=c_{f}} $ and with $ \mathrm{U^{BF}(5)} $
Hamiltonian if $\mathrm{c_{s}=0} $ and $\mathrm{c_{d}\neq c_{f}\neq0} $. So, the $\mathrm{c_{s}\neq
c_{d} \neq c_{f} \neq 0}$  situation just corresponds to $\mathrm{U^{BF}
(5)\leftrightarrow O^{BF}(6)}$ transitional region.  Hamiltonian Eq.(21) in even-even nuclei is equivalent to $ \mathrm{O(6)}$ Hamiltonian when $\mathrm{c_{s}=c_{d}}$ and with $ \mathrm{U(5)} $
Hamiltonian if $\mathrm{c_{s}=0 }$ and $\mathrm{c_{d}\neq0} $ and Hamiltonian in transitional region with $\mathrm{c_{s}\neq
c_{d} \neq 0}$.
 In our
calculation, we take  $\mathrm{c_{d}(=1)}$ constant value and $\mathrm{c_{s}}$
and  $\mathrm{c_{f}}$ change between 0 and $\mathrm{c_{d}}$.

For evaluating the eigenvalues of
Hamiltonian Eq.(21) the eigenstates are considered as
\begin{equation}
|k;\nu_{s}\nu n_\Delta LM \rangle= \textit{N}
S_{BF}^{+}(x_{1})S_{BF}^{+}(x_{2})S_{BF}^{+}(x_{3})...S_{BF}^{+}(x_{k})|lw\rangle^{BF}
\end{equation}
\begin{equation}
S _{x_{i} }^{+}=\frac {c_{s}}{1-c_{s}^{2} x_{i} } S_{B}^{+} (s)+\frac
{c_{d}}{1-c_{d}^{2} x_{i} } S_{B}^{+} (d)+\frac
{c_{f}}{1-c_{f}^{2} x_{i} } S_{F}^{+} (f)
\end{equation}
The lowest weight state, $ \mathrm{|lw\rangle^{BF}} $ , is defined as
\begin{align}
|lw\rangle^{BF}&=|\mathcal{N}=N_{B}+N_{F},k_{d}=\frac{1}{2}
(\nu_{d}+\frac{5}{2}),\mu_{d}=\frac{1}{2}
(n_{d}+\frac{5}{2}),k_{s}=\frac{1}{2}
(\nu_{s}+\frac{1}{2}),\mu_{s}=\frac{1}{2} (n_{s}+\frac{1}{2}), k_{f}=\frac{1}{2}
\nonumber \\
&(\nu_{f}-\frac{2j+1}{2}),\mu_{f}=\frac{1}{2} (n_{f}-\frac{2j+1}{2}),J ,M \rangle
\end{align}
\begin{equation}
S_{n}^{0} |lw\rangle^{BF} =\Lambda_{n}^{0} |lw\rangle^{BF} \quad , \quad \Lambda_{n}^{0} = c_{s}^{2n}(\nu_{s}+\frac {1}{2})\frac
{1}{2}+c_{d}^{2n}(\nu_{d}+\frac {5}{2})\frac {1}{2}+c_{f}^{2n}(\nu_{f}-\frac {2j+1}{2})\frac {1}{2}
\end{equation}
The eigenvalues of Hamiltonian Eq.(21) can then be expressed as
\begin{equation}
E^{(k) }=h^{(k) }+\alpha \Lambda_{1}^{0}+\beta
(\tau _{1}(\tau _{1}+3)+\tau _{2}(\tau _{2}+1))+\gamma J(J+1)\quad  , \quad h^{(k) }=\sum_{i=1}^{k}{\frac {\alpha}{x_{i}}}
\end{equation}
In order to obtain the numerical results for energy spectra
$\mathrm{(E^{(k) } )}$ of the considered nuclei, a set of non-linear
Bethe-ansatz equations (BAE) with k- unknowns for k-pair
excitations must be solved. Also constants of
Hamiltonian with the least square fitting processes to experimental
data are obtained. To achieve this aim, we have changed variables as
$$ \mathrm{C_{s}=\frac  {c_{s}}{c_{d}} \leq 1    ,   C_{f}=\frac  {c_{f}}{c_{d}} \leq 1        ,    y_{i}=c_{d}^{2} x_{i}} $$
\begin{equation}
\frac {\alpha}{y_{i}}=\frac{ C_{s}^{2} (\nu_{s}+\frac
{1}{2})}{1-C_{s}^{2} y_{i}}+\frac{ (\nu_{d}+\frac
{5}{2})}{1- y_{i}}+\frac{C_{f}^{2} (\nu_{f}-\frac {2j+1}{2})}{1-C_{f}^{2} y_{i}}-{\sum_{j\neq i}{\frac {2}{y_{i}-y_{j}}}}
\end{equation}
The quantum number $\mathrm{(k)}$ is related to  $\mathrm{\mathcal{N}}$
by
$\mathrm{ \mathcal{N}=2k+\nu_{s}+\nu_{d}+\nu_{f}}$.
 The quality of the fits is specified by the values of $ \mathrm{\sigma = (\frac {1}{N_{tot}}\sum _{i,tot}{|E_{exp}(i)-E_{Cal}(i)|^2})^{\frac {1}{2}}}$(keV)
and $ \mathrm{\phi = \frac{\sum\limits_{i}|E_{i}^{theor}-E_{i}^ {exp}| } {\sum\limits_{i} E_{i}^{exp}} }$ (\%)
($\mathrm{N_{tot}}$ the number of energy levels where included in the
fitting processes)\cite{4,15}.

The complete study of the properties of quantum phase transitions comprises both the classical and  quantal analyses.
In this study, we focus  only on the quantal analysis and present the calculated phase transition
observables such as the level crossing, the expectation value of the d-boson
number operator and the expectation value of the fermion number operator.

 Once the eigenvalues have been obtained,  we can display how the energy levels change within the whole range of the $ \mathrm{C_{s}} $ and $ \mathrm{C_{f}}$ control parameters.  Fig.\ref{fig:2}  shows the energy surfaces of Hamiltonian of Eq.(21) for the  neighboring even-even (left panel) and odd-A nuclei (right
 panel). The calculations are performed by considering the same fit parameters for these nuclei,where the parameters are $ \mathrm{\alpha=1000} $keV, $\mathrm{\beta=-1.29} $keV,  $\mathrm{\gamma=6.05} $keV, N=10 . Fig.\ref{fig:2} shows how the
 energy levels as a function of the control parameter $ \mathrm{C_{s}} $ and $\mathrm{ C_{f}}$ evolve
 from one dynamical symmetry limit to the other . It can be seen
 from Figs that numerous level crossings occur. The crossings are due to the fact
 that $ \mathrm{\nu_{d}} $, $\mathrm{O(5)}$ quantum number called seniority, is
 preserved along the whole path between  O(6) and U(5)
 \cite{12,17}.

The other quantal order parameters that we consider here are the expectation values of the d-boson number operator and the expectation values of the fermion number operator.The expectation values of the d-boson number operator and fermion number operator are obtained as
\begin{equation}
\langle \hat{n_{d}}\rangle=\frac {\langle \psi |\hat{n_{d}}|\psi\rangle}{N}=\frac {2C_{s}^{2}C_{f}^{2}(\Lambda_{0}^{0}+k)-2(C_{s}^{2}+C_{f}^{2})(\Lambda_{1}^{0}+k y_{1}^{-1})+2(\Lambda_{2}^{0}+k y_{2}^{-2})}{N (1-C_{s}^{2})(1-C_{f}^{2})}-\frac {5}{2N}
\end{equation}
\begin{equation}
\langle \hat{n_{f}}\rangle=\frac {\langle \psi |\hat{n_{f}}|\psi\rangle}{N}=\frac{2(1+C_{s}^{2})(\Lambda_{0}^{0}-\Lambda_{1}^{0}+k(1-y_{1}^{-1}))-2(\Lambda_{0}^{0}-\Lambda_{2}^{0}+k(1-y_{2}^{-2}))}{N(1-C_{f}^{2})(C_{s}^{2}-C_{f}^{2})}+\frac {2j+1}{2N}
\end{equation}
Fig.\ref{fig:3} shows the expectation values of the d-boson number operator
for the lowest states even-even (left panel) and odd-A nuclei (right
panel) as a function of  control parameters for
N=10 bosons. Fig.\ref{fig:3} (left panel) displays that the expectation values of the
number of d bosons for each L, $\mathrm{n_{d}}$, remain approximately
constant for $ \mathrm{C_{s}  <0.45} $ and only begin to change rapidly for $ \mathrm{C_{s}
>0.45}$. The near constancy of $\mathrm{n_{d}}$ for $\mathrm{C_{s} <0.45}$, is an obvious
indication that  U(5) dynamical symmetry is preserved in this
region to a high degree and also the $\mathrm{n_{d}}$ values change
rapidly with $\mathrm{ C_{s}} $ over the range $\mathrm{0.65\leq C_{s} \leq 1}$.
Fig.\ref{fig:4} shows likewise $\mathrm{\langle n_{f}\rangle}$ as a function of the $ \mathrm{C_{s}} $ and $ \mathrm{C_{f}}$ control parameters.

Nuclear physics has made important
contributions to study QPTs because nuclei display a
variety of phases in systems ranging from few to many
particles \cite{18}. The nuclei in the mass regions around
130  have transitional characteristics, intermediate
between the spherical and gamma-unstable shapes\cite{19}. The calculations have been done along Xe and Ba isotopic chains exhibit that
$\mathrm{^{134}Ba}$ and $\mathrm{^{130}Xe}$ are the best candidates  E(5) critical point symmetry \cite{19,20}. The first Bose-Fermi critical symmetry, called E(5/4), has been proposed for an odd-A system in Ref.\cite{11}. A search for experimental examples of E(5/4) should concentrate on the case that a $j=3/2$ particle if coupled to an  E(5) core, $ \mathrm{^{135}Ba}$ and $\mathrm{^{131}Xe}$ built on the single particle neutron in the $ \mathrm{2d_{\frac{3}{2}}}$ shell model orbit should be possible candidates \cite{11,19,20}. In what follows we describe a simultaneous analysis of $\mathrm{^{134}Ba}$ with  $ \mathrm{^{135}Ba}$ and  $\mathrm{^{130}Xe}$ with  $\mathrm{^{131}Xe}$ within the  U(6/4) supergroup.
 In order to obtain energy spectrum and
realistic calculation for these nuclei, we need to specify
Hamiltonian parameters Eq.(21). According to the supersymmetry concept, the even-even and odd-A nuclei are described by the same set of fit parameters, thus to achieve a better fit the states of both nuclei were used. The best fits for Hamiltonian's parameters, namely $
\mathrm{\alpha}$, $ \mathrm{\beta} $ and  $\mathrm{\gamma}  $,  used in the
present work are shown in Table 1. A comparison between the available experimental levels and the
predictions of our results for the $ \mathrm{^{134-135}Ba}$
and  $ \mathrm{^{130-131}Xe}$
 isotopes in the low-lying region of spectra along with $\mathrm{R_{\frac{4}{2}}=\frac {E(\nu_{d}=2)}{E(\nu_{d}=1)})}$ values are shown in Fig. \ref{fig:5} and  Fig. \ref{fig:6}, respectively.The $\mathrm{R_{\frac{4}{2}}}$ value is one of the most basic structural predictions of $ \mathrm{U
 (5)-O(6)} $ transition\cite{19,20}. The ratio equal to  2.2-2.3
indicates the spectrum of transitional nuclei\cite{19,20}.
We
have tried to extract the best set of parameters which reproduce
these complete spectra with minimum variations. It means that our
suggestion to use this transitional Hamiltonian for the
description of the Ba and Xe isotopes  would not have any
contradiction with other theoretical studies done with special
hypotheses about mixing of intruder and normal configurations.
 So, we conclude
 from the values of control parameter which has been obtained and
 $\mathrm{R_{\frac{4}{2}}}$ value,  that $ \mathrm{_{56}^{134-135}Ba }$ and  $ \mathrm{_{54}^{130-131}Xe} $ isotopes are
 the best candidates for  U(5)-O(6) transition in  U(6/4)  supersymmetry scheme.

Although, studies of QPTs in odd-even and even-even nuclei were extensively done, our results are novel, since (1) we have proposed exactly-solvable  supersymmetry Richardson-Gaudin (R-G) model for transitional region by which we can be investigate the phase transition observables in the both of nuclei by using of the concept of supersymmetry (2) The experimental evidences have been presented for E(5) and E(5/4) nuclei and have been analyzed them by supersymmetry scheme.
The important new result of the present paper is to have employed the nuclear supersymmetry approach for description of the transitional region between spherical and  gamma -unstable phase shape in addition to dynamical symmetry limits in one chain isotopic.
\begin{table}
\begin{center}
\begin{tabular}{p{9.3cm}}

\footnotesize Table 1. \footnotesize Parameters of Hamiltonian
Eq.(20)  used in the calculation of the Ba and Xe nuclei.
\footnotesize All parameters  are given in keV.\\
\end{tabular}

\begin{tabular}{ccccccccc}

\hline
Nucleus      &$\mathcal{N}$    & $\mathrm{C_{s}}$ & $\mathrm{C_{f}}$ & $\mathrm{\alpha}$ &  $\mathrm{\beta}$ & $\mathrm{\gamma}$ & $\mathrm{\sigma}$  & $\mathrm{\phi}$  \\
\hline

\quad\\
$\mathrm{^{134}Ba-^{135}Ba}$ & 5 & 0.52 & 0.8 & 333.12 & 1.945 & 0.73 & 155.42 & 10.73\%  \\
\quad\\
$\mathrm{^{130}Xe-^{131}Xe}$ & 5 & 0.55 & 0.9 & 187.84 & 0.9564 & 10.98 & 143.29 & 12.4\% \\
 \quad\\

\hline
\end{tabular}
\end{center}
\end{table}

\clearpage
\begin{figure}
\begin{center}
\includegraphics[height=6cm]{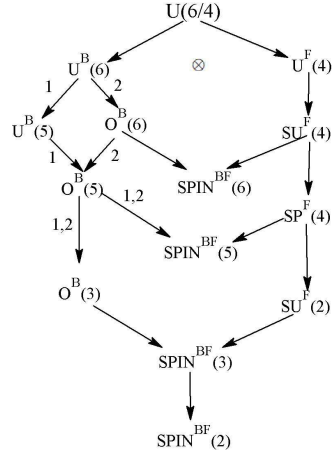}
\caption{The lattice of algebras in the U(6/4) supersymmetry scheme.\label{fig:1}}
\end{center}
\end{figure}
\begin{figure}
\includegraphics[height=4.5cm]{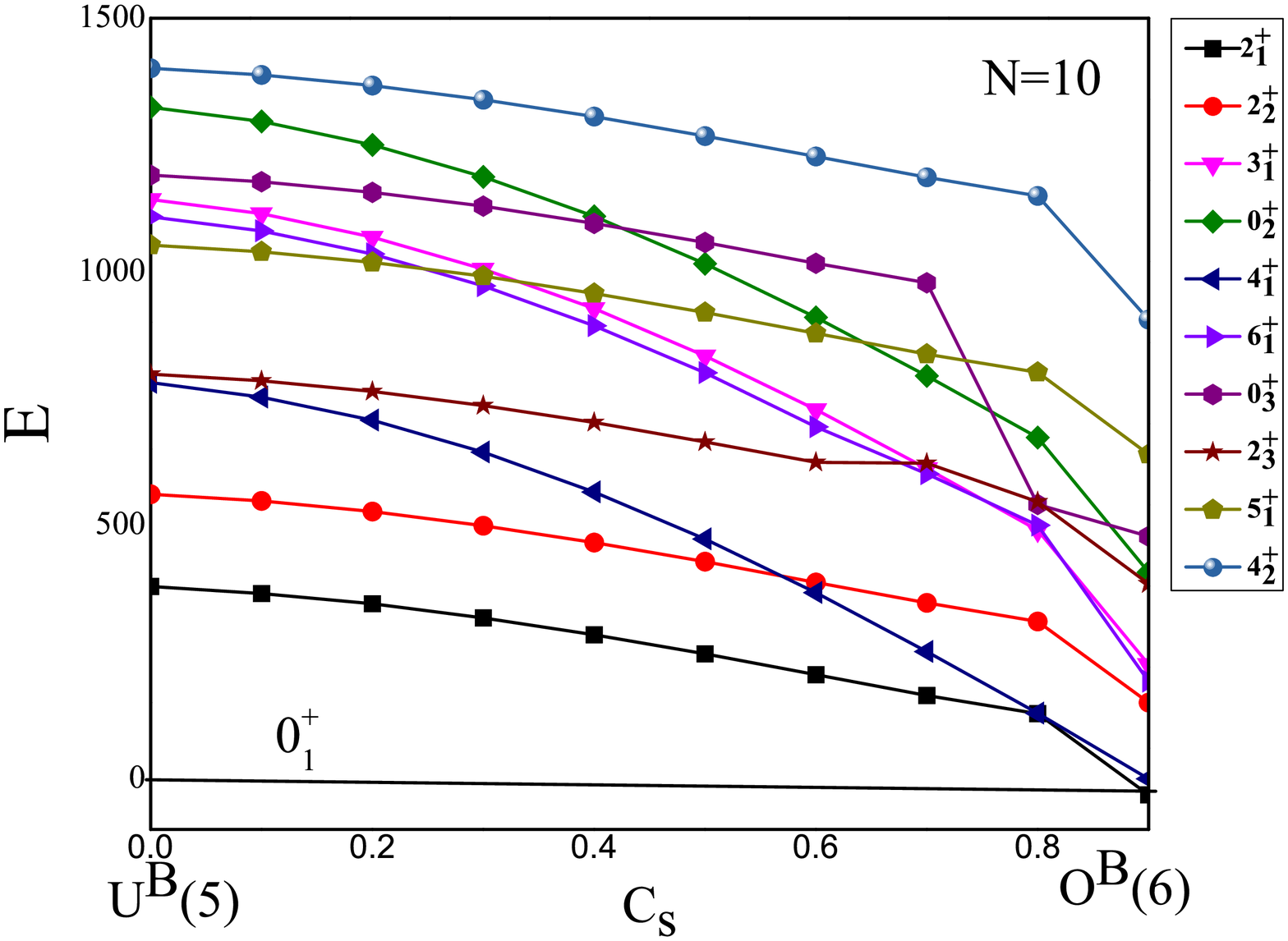}
\includegraphics[height=4cm]{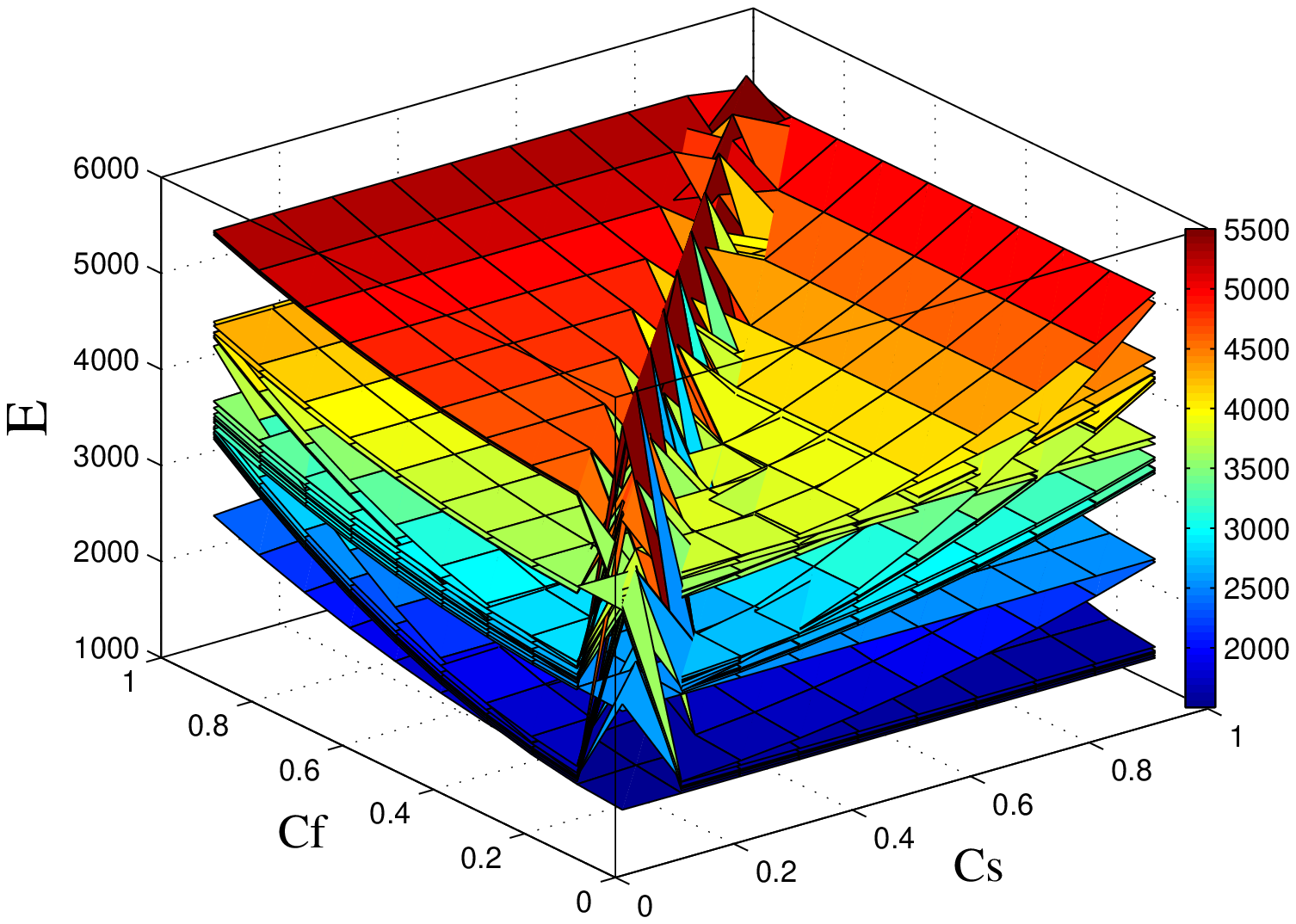}
\caption{Energy levels as a function of  $ \mathrm{C_{s}} $ control parameter for a even-even nuclei (left panel) and for odd-A nuclei as a function of the $ \mathrm{C_{s}} $ and $ \mathrm{C_{f}}$ control parameters(right
panel)
in the Hamiltonian (20) for N=10 bosons
with$\mathrm{\alpha=1000}$keV,$\mathrm{\beta=-1.29}$keV,$\mathrm{\gamma=6.05}$keV.\label{fig:2}}
\end{figure}
\begin{figure}
\includegraphics[height=4.5cm]{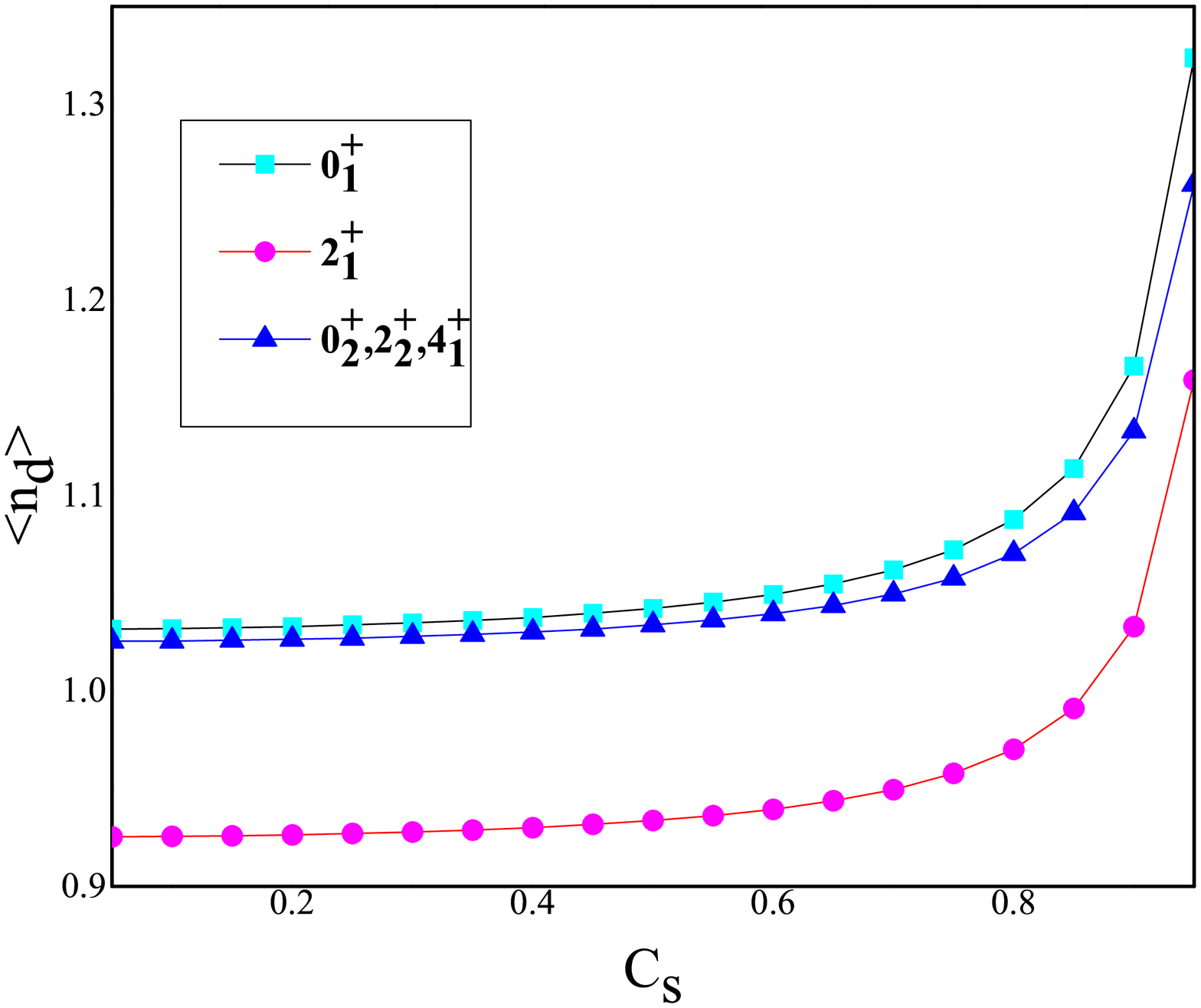}
\includegraphics[height=4cm]{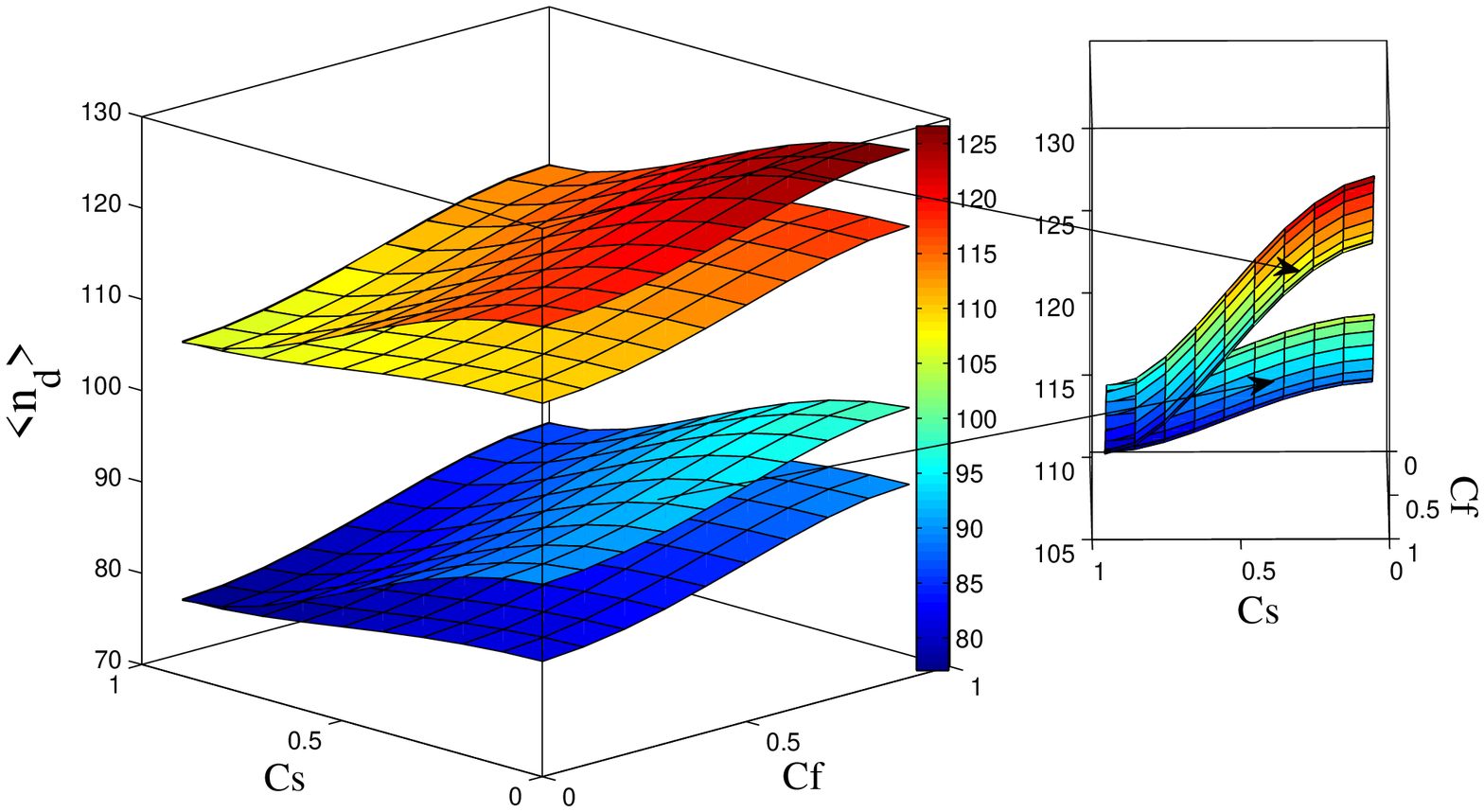}
\caption{The expectation values of the d-boson number operator
for the lowest states as a function of  $ \mathrm{C_{s}} $ control parameter for an even-even nuclei (left panel) and for odd-A nuclei as a function of the $ \mathrm{C_{s}} $ and $\mathrm{ C_{f}}$ control parameters(right
panel).\label{fig:3}}
\end{figure}
\begin{figure}
\includegraphics[height=6cm]{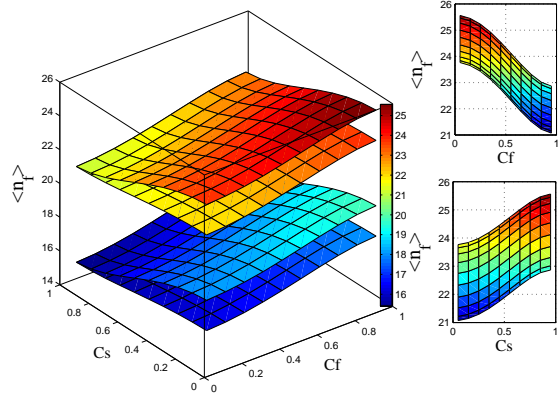}
\caption{The expectation values of the fermion number operator for odd-A nuclei for the lowest states as a function of  $\mathrm{ C_{s}} $ and $ \mathrm{C_{f}}$ control parameters.\label{fig:4}}
\end{figure}
\begin{figure}
\begin{center}
\includegraphics[height=5cm]{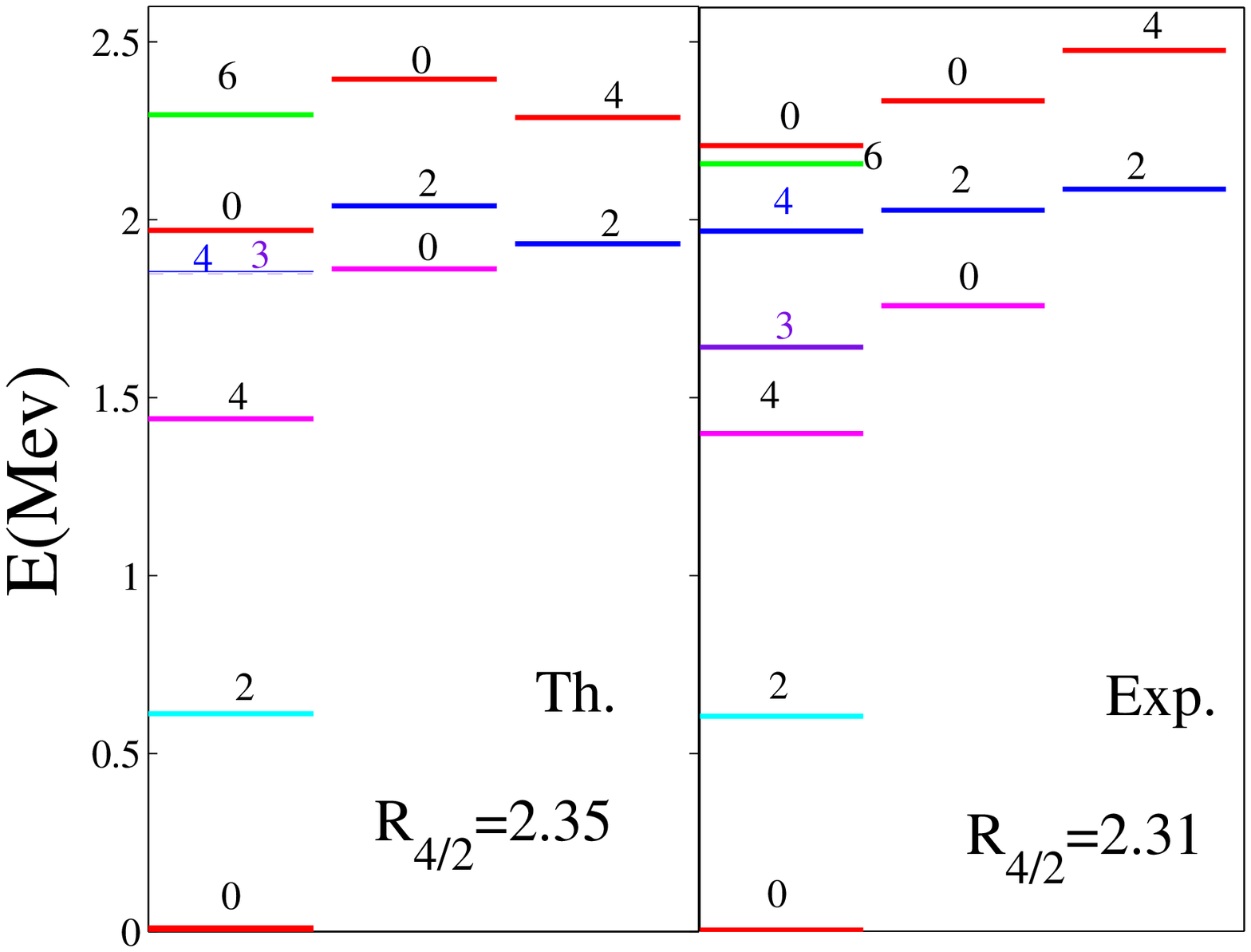}\includegraphics[height=5cm]{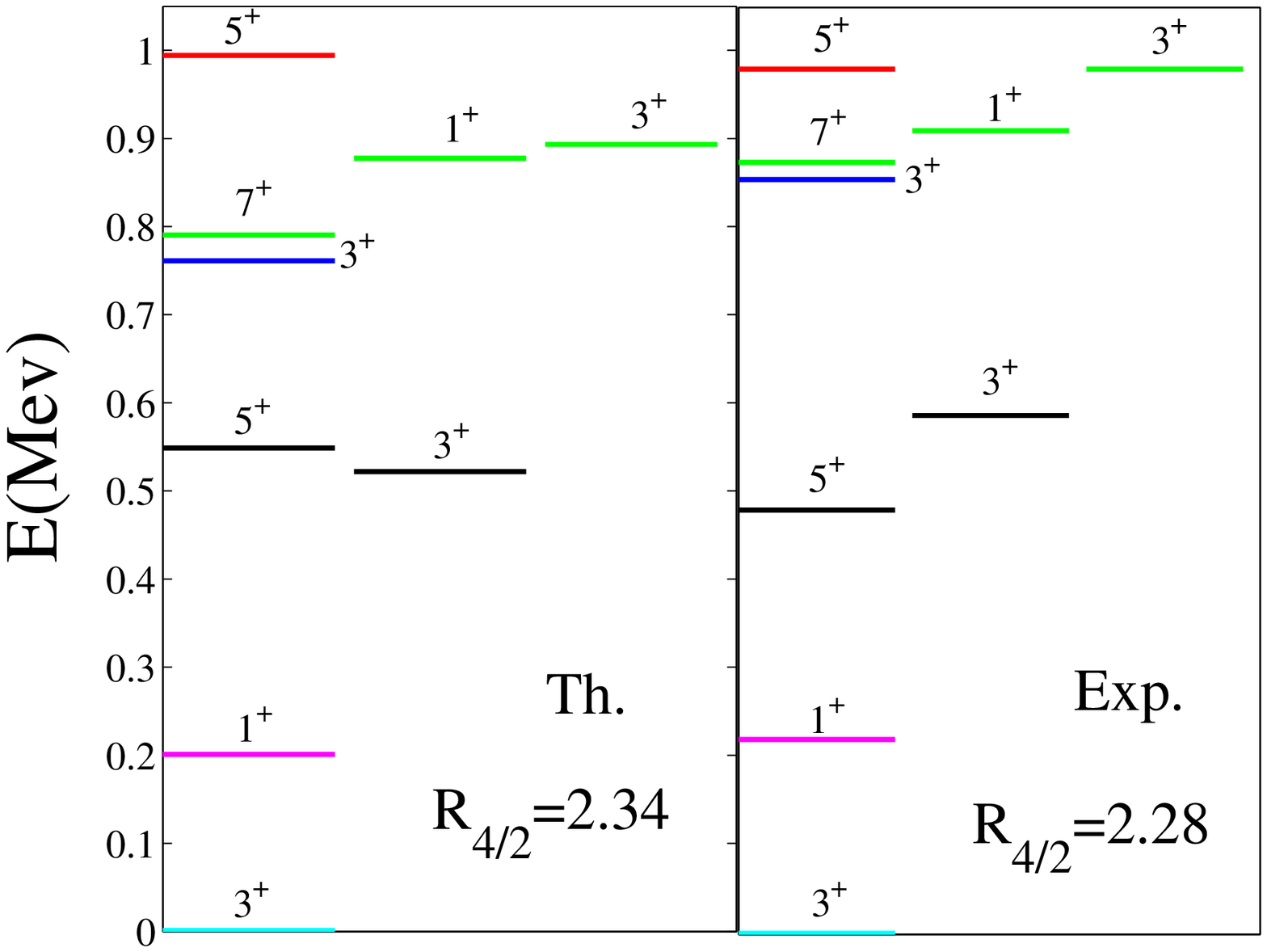}

\caption{Comparison between calculated and experimental spectra
of positive parity states in $\mathrm{^{134}Ba }$ (left panel) and $\mathrm{^{135}Ba} $(right panel) . The parameters of the
calculation are given in Table 1. The experimental spectra,
is taken from.\cite{21}\label{fig:5}}
\end{center}
\end{figure}
\begin{figure}
\begin{center}
\includegraphics[height=5cm]{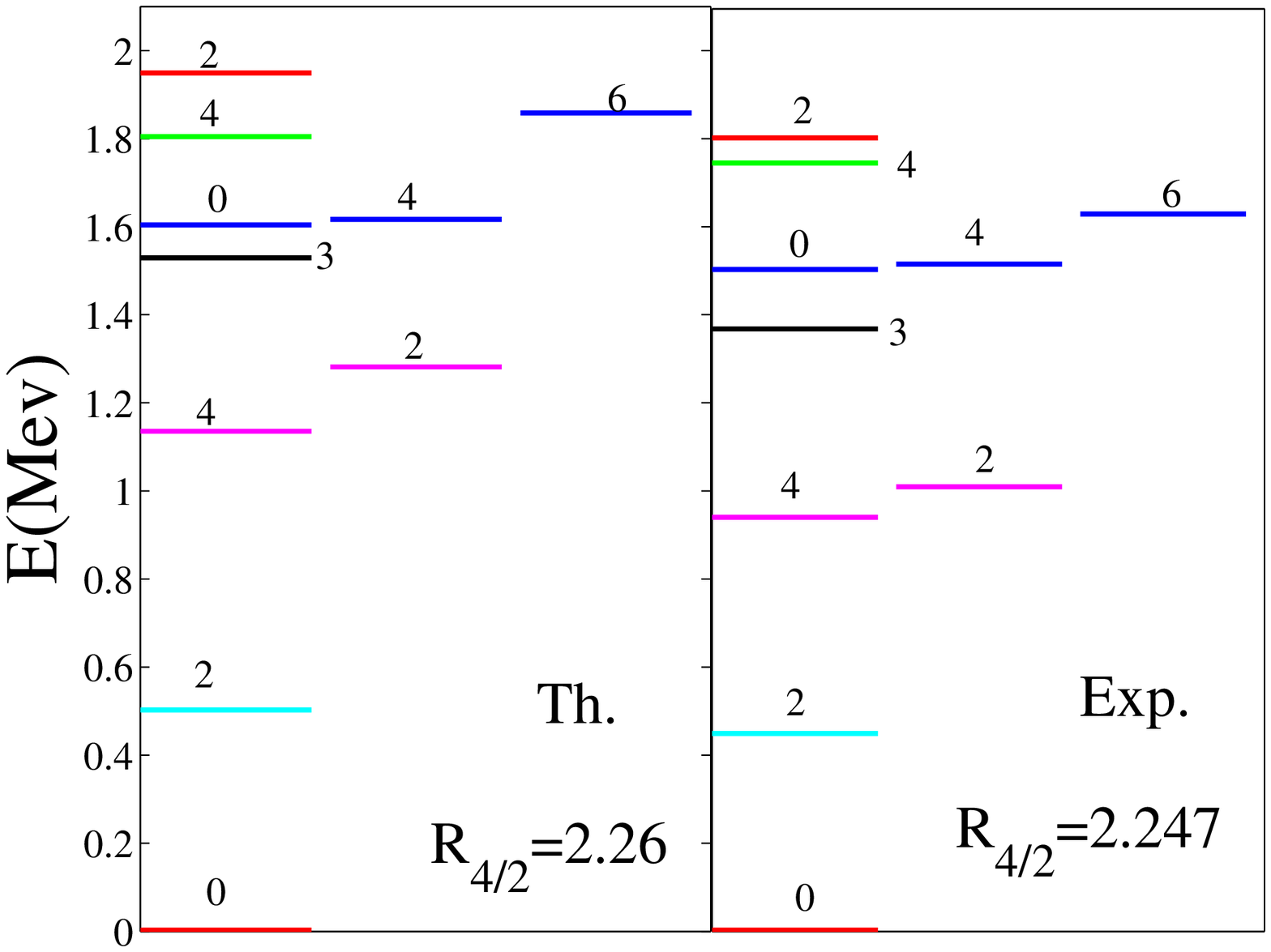}\includegraphics[height=5cm]{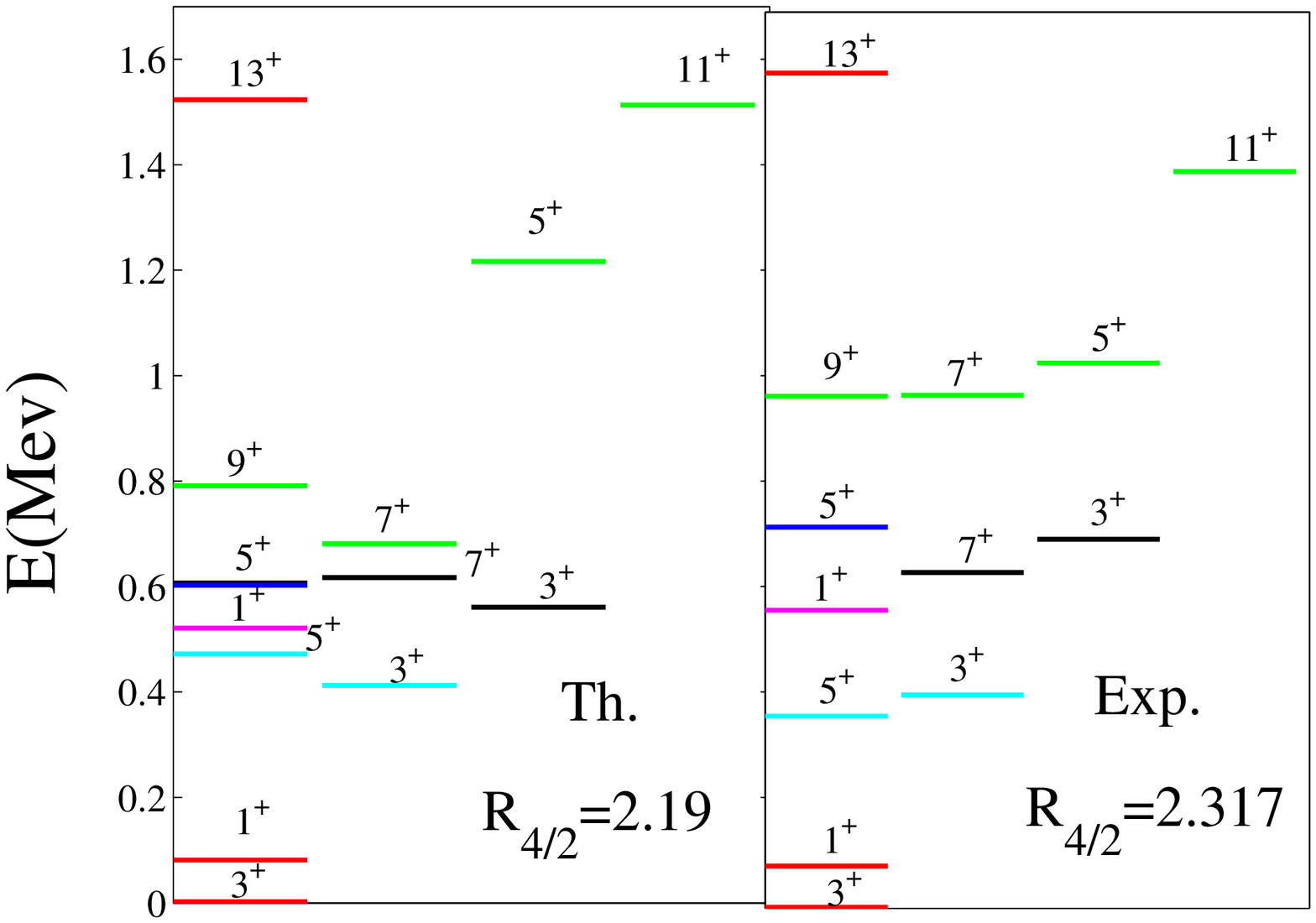}

\caption{Comparison between calculated and experimental spectra
of positive parity states in $\mathrm{^{130}Xe }$(top panel) and  $\mathrm{^{131}Xe} $(bottom panel). The parameters of the
calculation are given in Tables 1. The experimental spectra,
is taken from.\cite{21}\label{fig:6}}
\end{center}
\end{figure}

\end{document}